# Free Fermion Antibunching in a Degenerate Atomic Fermi Gas Released from an Optical Lattice

T. Rom[1], Th. Best[1], D. van Oosten[1], U. Schneider[1], S. Fölling[1], B. Paredes[1], and I. Bloch[1]

[1]*Institut für Physik, Johannes Gutenberg-Universität, Staudingerweg 7, 55128 Mainz, Germany*

**Noise in a quantum system is fundamentally governed by the statistics and the many-body state of the underlying particles[1-4]. Whereas for bosonic particles the correlated noise[5-7] observed for e.g. photons[8] or bosonic neutral atoms[9-14] can still be explained within a classical field description with fluctuating phases, the anticorrelations in the detection of fermionic particles have no classical analogue. The observation of such fermionic antibunching is so far scarce and has been confined to electrons[15-17] and neutrons[18]. Here we report on the first direct observation of antibunching of neutral fermionic atoms. Through an analysis of the atomic shot noise[3,10,19] in a set of standard absorption images, of a gas of fermionic $^{40}$K atoms released from an optical lattice, we find reduced correlations for distances related to the original spacing of the trapped atoms. The detection of such quantum statistical correlations has allowed us to characterise the ordering and temperature of the Fermi gas in the lattice. Moreover, our findings are an important step towards revealing fundamental fermionic many-body quantum phases in periodic potentials, which are at the focus of current research.**

Bunching and antibunching effects in the detection of bosonic and fermionic particles are usually understood as a consequence of the constructive or destructive interference of the two possible propagation paths that two particles can follow to reach two detectors, as was first outlined in the famous experiments of Hanbury-Brown and Twiss for the case of bosonic particles[7]. For the case of a fermionic atom cloud released from a periodic potential, the origin of spatial anticorrelations in the shot noise can be more naturally understood within the following picture. In the lowest energy band of a one-dimensional periodic potential, atoms can occupy Bloch states, each of them characterised by a crystal momentum $\hbar q$, as illustrated in Fig. 1. Each Bloch state is a superposition of plane waves, which correspond to different real momenta $p_\nu = \hbar q + \nu \cdot 2\hbar k$, where $k = 2\pi/\lambda$, $\lambda$ being the wavelength of the laser light used to form the optical lattice potential, and $\nu$ an integer number. When a particle with crystal momentum $\hbar q$ is released from the lattice potential, its wave function will expand freely, and after a time of flight (TOF) $t$, long enough to neglect the initial size of the system, it can be detected at any of the positions $x_\nu = (\hbar q + \nu \cdot 2\hbar k)t/m$. These positions are equally spaced by $\ell = 2\hbar k t/m$, where $m$ is the atomic mass. Conversely, if a particle is detected at any of the positions $x_\nu$, it had to emerge from a state with crystal momentum $\hbar q$. Since the Pauli principle does not allow two fermionic particles to occupy the same Bloch state, the output of two detectors at positions $x_\nu$ and $x_{\nu'}$, will thus be anticorrelated for any $\nu,\nu'$. That is, if one detector detects a particle, the second detector will not detect another particle. Therefore, if $n(x)$ represents the density of the expanding atom cloud detected at position $x$ in a single run of the experiment, the spatial correlations $\langle n(x)n(x')\rangle$ will vanish for any $|x-x'|$ being an integer multiple of $\ell$. It is interesting to note that for bosonic particles, which can populate the same crystal momentum state, enhanced spatial correlations at these distances can arise, corresponding to a "bunching" of the particles[10,14]. For both bosons and fermions, the behaviour of $\langle n(x)n(x')\rangle$ for $|x-x'|$ different from these characteristic distances will depend on the many-body state of the trapped particles[3].

The starting point for our experiment is an oblate shaped degenerate Fermi gas in a crossed optical dipole trap with typically $2.0(5)\times10^5$ $^{40}$K atoms at temperatures as low as $T/T_F$=0.23(3). These atoms are transferred into a three-dimensional optical lattice potential with lattice depths of $V_x=V_y=20$ $E_r$ and $V_z=10$ $E_r$ in the $x,y,$ and $z$-direction, respectively. Here, $E_r= \hbar^2 k^2/2m$ denotes the recoil energy. For our experimental parameters the atoms form a fermionic band insulator in the lowest energy band, which we verify using an adiabatic mapping technique[20,21] (see inset Fig. 2a). For the correlation analysis, all optical potentials are suddenly switched off and images are recorded along the vertical direction ($z$) via standard absorption imaging after a TOF period of 10 ms.

Typically, 200 absorption images are taken for the same set of experimental parameters, yielding independent two-dimensional spatial density profiles $n(\mathbf{x})$. An example of such a density profile and a corresponding cut can be seen in Figs. 2a,b. The atom cloud exhibits a Gaussian like envelope, as a result of the expansion of the Wannier function that characterises the initial localisation of the particles at each lattice site. Additionally, atomic noise is directly visible as fluctuations in the recorded single images. However, no distinct structure can be seen in this noise.

After discarding images showing technical artefacts (approx. 30% of the total images), a spatial correlation analysis is carried out. This yields the density-density correlation function

$$C(\mathbf{d}) = \frac{\int \langle n(\mathbf{x}-\mathbf{d}/2)\, n(\mathbf{x}+\mathbf{d}/2)\rangle\, d^2x}{\int \langle n(\mathbf{x}-\mathbf{d}/2)\rangle \langle n(\mathbf{x}+\mathbf{d}/2)\rangle\, d^2x}, \qquad (1)$$

which gives the conditional probability of finding particles at two positions separated by a vector $\mathbf{d}$, averaged over the whole image. The brackets $\langle\cdot\rangle$ denote a statistical averaging over the set of images. Values of $C(\mathbf{d}) > 1$ correspond to bunching, whereas the characteristic antibunching for fermions shows up as $C(\mathbf{d}) < 1$.

Results of the above analysis are shown in Figs. 2c,d. Centred around the autocorrelation peak, a perfectly regular structure arises (see Fig. 2c) in the correlation images. A horizontal cut through the centre of such a correlation image exhibits distinct dips where $C(\mathbf{d}_{\text{dip}}) < 1$ (see Fig. 2d), demonstrating the antibunching of the atoms. We find that the spacing between these dips along the horizontal and vertical direction is in good agreement with the expected value of $\ell$. In order to reliably detect and characterise these correlation dips, the signal to noise ratio in the resulting correlation image has to be sufficiently high. As the root mean square (RMS) background noise decreases in good agreement with a $1/\sqrt{N}$ scaling, the signal to noise ratio can be successively increased by including more single shot absorption images in the correlation analysis. When fitting two-dimensional gaussians to these dips, we find that at least 50 images are needed for a reliable fit, if width, position and amplitude of the dips are free parameters.

In order to experimentally investigate the temperature dependence of the measured correlation signal, we have prepared Fermi gases at different temperatures by varying the end point of the forced evaporation procedure in the optical dipole trap. Before the lattice potentials are ramped up, the optical trap depth is adjusted to obtain a defined external confinement, equal for all temperatures. For each temperature, a correlation analysis for a series of absorption images is performed. As Fig. 3 shows, we find a distinct decrease of the correlation amplitude with increasing temperature. Antibunching is, however, still clearly visible up to temperatures of $T/T_F \approx 1$. Note that all quoted temperatures refer to the initial temperatures in the optical dipole trap without the lattice, as we currently do not have an independent means to measure the temperature of the quantum gas in the lattice.

In order to understand the strength of the correlation signals and specifically their temperature dependence, we consider a one-dimensional system of $N$ fermionic atoms in a lattice in the presence of an external harmonic confinement. We assume that the lattice is deep enough to suppress tunnelling processes and that the temperature is low enough such that the atoms only occupy the lowest onsite vibrational state. Under these conditions, the occupation of the lattice sites is given by a Fermi-Dirac distribution of the form:

$$f_j = \frac{1}{1+\exp\left(\left(\frac{1}{2}m\omega^2(\lambda/2)^2 j^2 - \mu\right)\Big/k_B T^{lat}\right)}, \qquad (2)$$

where the integer number $j$ labels the lattice site and $T^{lat}$ denotes the temperature in the lattice. The trapping frequency of the external harmonic confinement is denoted by $\omega$ and the chemical potential $\mu$ is defined through $\Sigma_j f_j = N$. At zero temperature, starting from the centre of the trap, the atoms fill the lattice sites one-by-one, forming a Fermi sea in real space (see inset Fig. 4). In analogy to the Fermi momentum of free fermions, we define a Fermi radius $R_F = N\lambda/4$, which characterises the extension of the atomic cloud. The Fermi temperature in the lattice $T_F^{lat}$ can then be written as $k_B T_F^{lat} = 1/2\, m\omega^2 R_F^2$. As temperature increases, atoms can access lattice sites further out in the trap, leading to an increase in the spatial extension of the atom cloud.

A similar calculation to the one presented in ref. [10] yields the fermionic noise correlator

$$C(d) = 1 - \frac{1}{N^2}\left|\sum_j f_j\, e^{i 2\pi d \cdot j/\ell}\right|^2, \qquad (3)$$

where the minus sign on the right-hand side causes dips at integer multiples of $\ell$. For zero temperature, the above equation takes the simple form $C(d) = 1 - \left[\dfrac{\sin(\pi d/\ell N)}{N \sin(\pi d/\ell)}\right]^2$. The temperature dependence of the shape of one of these dips is shown in Fig. 4. The amplitude of the dip is always equal to 1, regardless of temperature. However, its width decreases with increasing temperature. This leads to a decrease in the signal strength of the dip, which we characterise by the spatial integral of the correlation function (see eq. 3) over a distance $\ell$. We can understand these features in the following way. The amplitude of 1, corresponding to the vanishing of the correlation function $C(d)$, is a universal feature for fermions. It reflects the fact that two fermions cannot occupy the same crystal momentum state. On the other hand, the detailed shape of the dip directly reflects the density-density correlations in crystal momentum space of the fermionic system before expansion. These correlations can be characterised by a correlation length in momentum space, which determines the width of the dip. For non-interacting fermions, this correlation length is inversely proportional to the size of the system. This explains that when temperature increases, leading to an expansion of the system in the trap, the width of the dip decreases. It is interesting to note that for the case of zero temperature the wings of the dip oscillate with a period inversely proportional to the Fermi radius. These oscillations are analogous to the Friedel oscillations that a system of free fermions exhibits in real space.

Let us compare these findings to our experimental results. For a typical system size of $N=100$ in the $x$ and $y$-dimension, we expect the width of each dip to be on the order of 1 μm based on the above model, much smaller than the actual optical resolution of our imaging system of 5.6 µm. Therefore, the observed shape and width of our correlation dips is predominantly determined by the point spread function of our optical imaging system (approximately a gaussian) and does not allow us to observe the detailed shape of the correlation dips predicted by theory. However, as the point spread function leaves the volume of a correlation dip constant, a change in the width of the theoretically predicted signal directly translates into a change of the observed correlation amplitude. For our lowest temperatures, we observe a correlation amplitude of $8(1)\times10^{-4}$, in good agreement with the expected value at zero temperature of $9(4)\times10^{-4}$ for our experimental parameters, taking additionally into account the integration along the line of sight[10]. For increasing temperatures, we have compared the observed decrease in the correlation signal with the one predicted by a two-dimensional version of the model presented above (see methods). Assuming an adiabatic heating when the atoms are loaded into the lattice potential[22], we find good quantitative agreement with theory (see Fig. 3). These results show that the correlation signal can be used as an efficient thermometer for fermions in an optical lattice.

In conclusion, we have demonstrated antibunching of free neutral fermionic atoms using a degenerate single component Fermi gas released from a three dimensional optical lattice potential. The anticorrelations obtained from a shot noise analysis of standard absorption images have allowed us to reveal the quantum statistics and furthermore identify the ordering and temperature of the atoms in the periodic potential up to a temperature of $T/T_F \approx 1$. These measurements show that noise correlations are a robust tool for further studies of degenerate Fermi systems. Especially, this method could allow the unambiguous detection of fermionic quantum phases such as an antiferromagnetically ordered Néel phase[23], which is strongly connected to models of High-$T_c$ superconductivity[24,25]. Further complex orders could be revealed by applying this method to low dimensional quantum systems[4,26], mixtures[27,28], quantum phases with disorder[29] or supersolid phases[30].

**Note:** Recently it has come to our attention that fermionic antibunching has also been observed in an ultracold gas of metastable $^3$He by a collaboration between groups at Orsay and Amsterdam.

## Methods

**Experimental setup and cooling cycle.** In the experiment, we sympathetically cool $^{40}$K with $^{87}$Rb in a two step process. First, both species are trapped in a magnetic trap of Quadrupole-Ioffe type, $^{87}$Rb in the $|F=2, m_F=2\rangle$ and $^{40}$K in the $|F=9/2, m_F=9/2\rangle$ substates. Radio-frequency (rf)-evaporative cooling of $^{87}$Rb is used to cool the mixture to a temperature of 2 μK. Both species are then transferred into a crossed optical dipole trap formed at the intersection of two orthogonal elliptical gaussian laser beams ($w_{x,y}$= 150 μm, $w_z$=40 μm) at a wavelength of 1030 nm. After switching off the magnetic trap, a constant offset field of 13.6 G is applied. Then $^{87}$Rb is transferred to the $|F=1, m_F=1\rangle$ state with an adiabatic microwave sweep, and $^{40}$K is transferred to the $|F=9/2, m_F=-9/2\rangle$ state with an adiabatic rf-sweep across the magnetic sub-states. The depth of the optical dipole trap is subsequently lowered to further evaporate the two species. At the end of this cooling process, we are left with up to 2.2(5)×10$^5$ $^{40}$K and 2.0(5)×10$^5$ $^{87}$Rb atoms in the optical trap. After removing the rubidium cloud completely via a resonant laser pulse, we are left with a single component Fermi gas of typically 2.0(5)×10$^5$ atoms at an initial temperature $T$ as low as $T/T_F \approx 0.23(3)$.

**Optical lattices.** The cold fermionic quantum gas is loaded into a three-dimensional optical lattice potential formed by three mutually orthogonal optical standing waves. The laser beams used for the optical standing waves have waists of 150 μm at the position of the trapped quantum gas and a wavelength of $\lambda$=755 nm, blue detuned to the atomic $D_1$ and $D_2$ transitions of $^{40}$K. The lattice potentials are successively ramped up to final lattice depths of $V_x=V_y=20\,E_r$, and $V_z=10\,E_r$. This is done in an $S$-shaped ramp within 40 ms, first in the $x$ and $y$-directions and finally in the $z$-direction.

**Determining correlation signal amplitudes.** Correlation signal amplitudes are extracted from the unprocessed correlation images by fitting the dips of the correlation images with an inverted two dimensional gaussian. First, the positions and widths of the anticorrelation dips are determined from three correlation images with $T/T_F = 0.23(3)$. In subsequent fits to all images, the dip amplitudes were obtained using these widths and positions. The correlation signal amplitude of an image is given as the average over the four dips at positions corresponding to the principal reciprocal lattice vectors. Note that the central autocorrelation peak shows a substructure due to technical artefacts.

**Temperature dependence in two dimensions.** A generalization of the theory model described above to the two-dimensional case is straightforward. The correlation function takes then the form $C(\mathbf{d}) = 1 - \frac{1}{N^2}\left|\sum_{\mathbf{j}} f_{\mathbf{j}}\, e^{i2\pi \mathbf{j}\cdot \mathbf{d}/\ell}\right|^2$, where $\mathbf{j}=(j_x, j_y)$, with $j_x, j_y$ being integer numbers. The integral of the function C(**d**) over a $\ell \times \ell$ cell, which characterises the signal strength of the dips, can be calculated to be

$$\frac{\ell^2}{N^2}\sum_{\mathbf{j}} f_{\mathbf{j}}^2 = \frac{\ell^2}{N}\left(1 - \frac{T}{T_F^{lat}}\left(1 - e^{-T_F^{lat}/T}\right)\right), \qquad (4)$$

where $k_B T_F^{lat} = m\omega^2(\lambda/2)^2 N/2\pi$ is the Fermi temperature of the atoms in the two-dimensional lattice. This result has been used to determine the region of amplitudes that, taking into account the uncertainty in the number of particles and the heating of the atoms when loading the optical lattice, is expected for the experimental dips.

12

**Acknowledgements** We would like to thank the DFG and the EU under a Marie-Curie Excellence grant (QUASICOMBS) and IP (SCALA) for financial support. We acknowledge technical assistance of Thomas Berg in the construction of the apparatus.

**Author Information** The authors declare no competing financial interests. Correspondence and request for materials should be addressed to I. Bloch (bloch@uni-mainz.de).

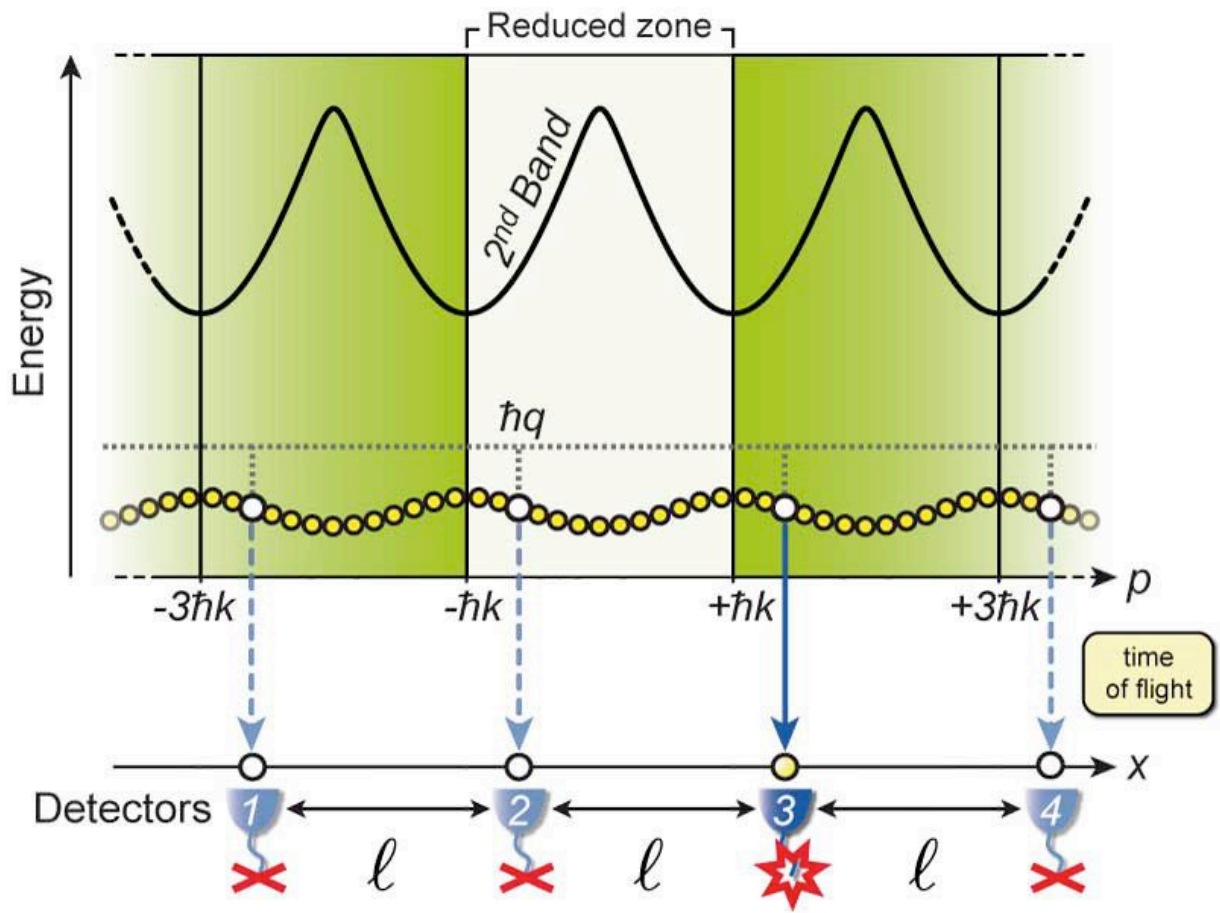

**Figure 1 | Origin of anticorrelations in a Fermi gas released from an optical lattice.** Each Bloch state, labelled by a crystal momentum $\hbar q$, is represented by a dot in the reduced zone scheme. The periodically extended zone scheme shows that each Bloch state is a superposition of states with momenta equally spaced by $2\hbar k$. After the lattice is switched off abruptly, an atom with crystal momentum $\hbar q$ propagates freely during a time of flight $t$ and can reach detectors equally spaced by a distance $\ell$. If e.g. detector number 3 detects a particle, then due to the single occupancy of each Bloch state dictated by the Pauli principle, detectors 1, 2 and 4 will not detect a particle.

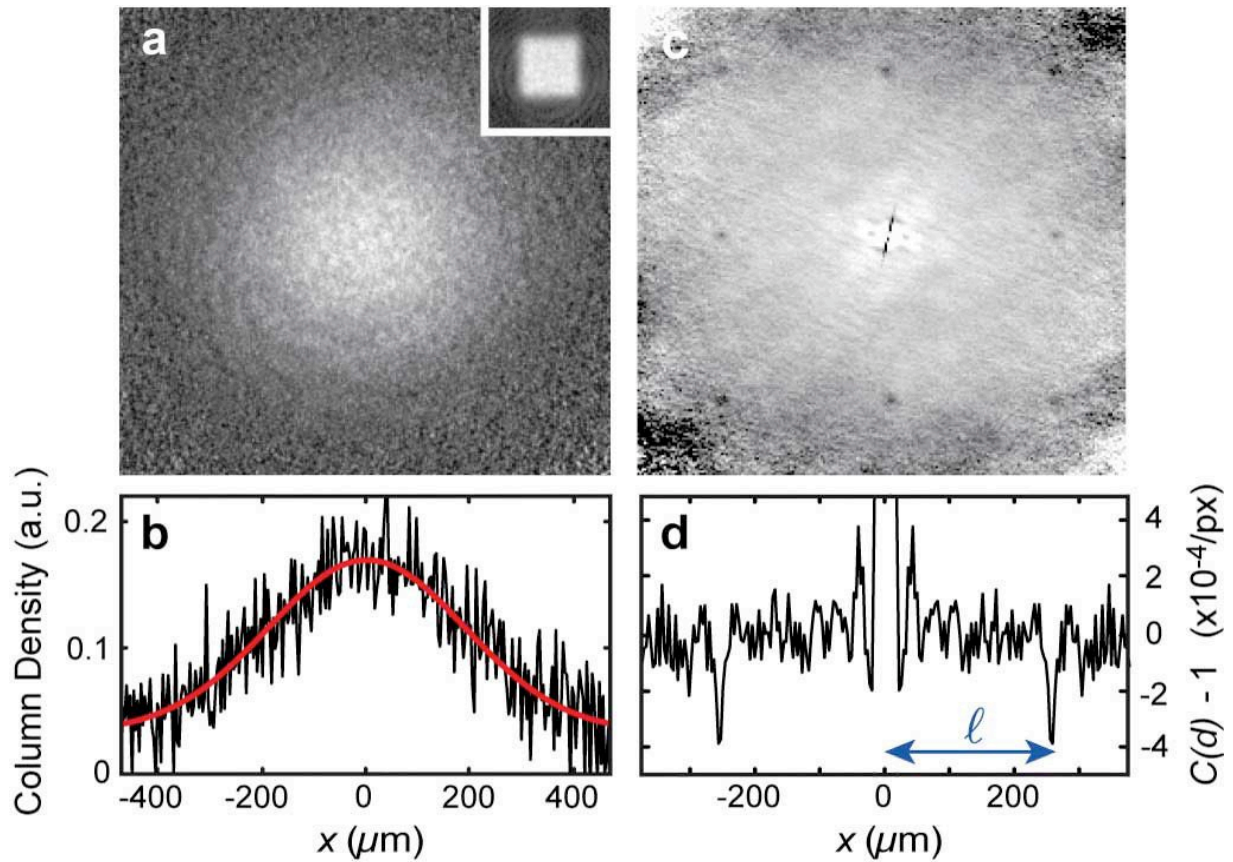

**Figure 2 | Single shot absorption images and correlation analysis. (a)** Single absorption image of a fermionic $^{40}$K atom cloud after 10 ms of free expansion. The inset shows a Brillouin zone mapping of the cloud, demonstrating that the Fermi gas is in a band insulating state. **(b)** 1D cut through the same picture together with a Gaussian fit (red). **(c)** Spatial noise correlations obtained from an analysis of 158 independent images, showing an array of eight dips. **(d)** Horizontal profile through the centre of the correlation image. The profile has been high-pass filtered to suppress a broad Gaussian background that we attribute to shot to shot fluctuations in the atom number.

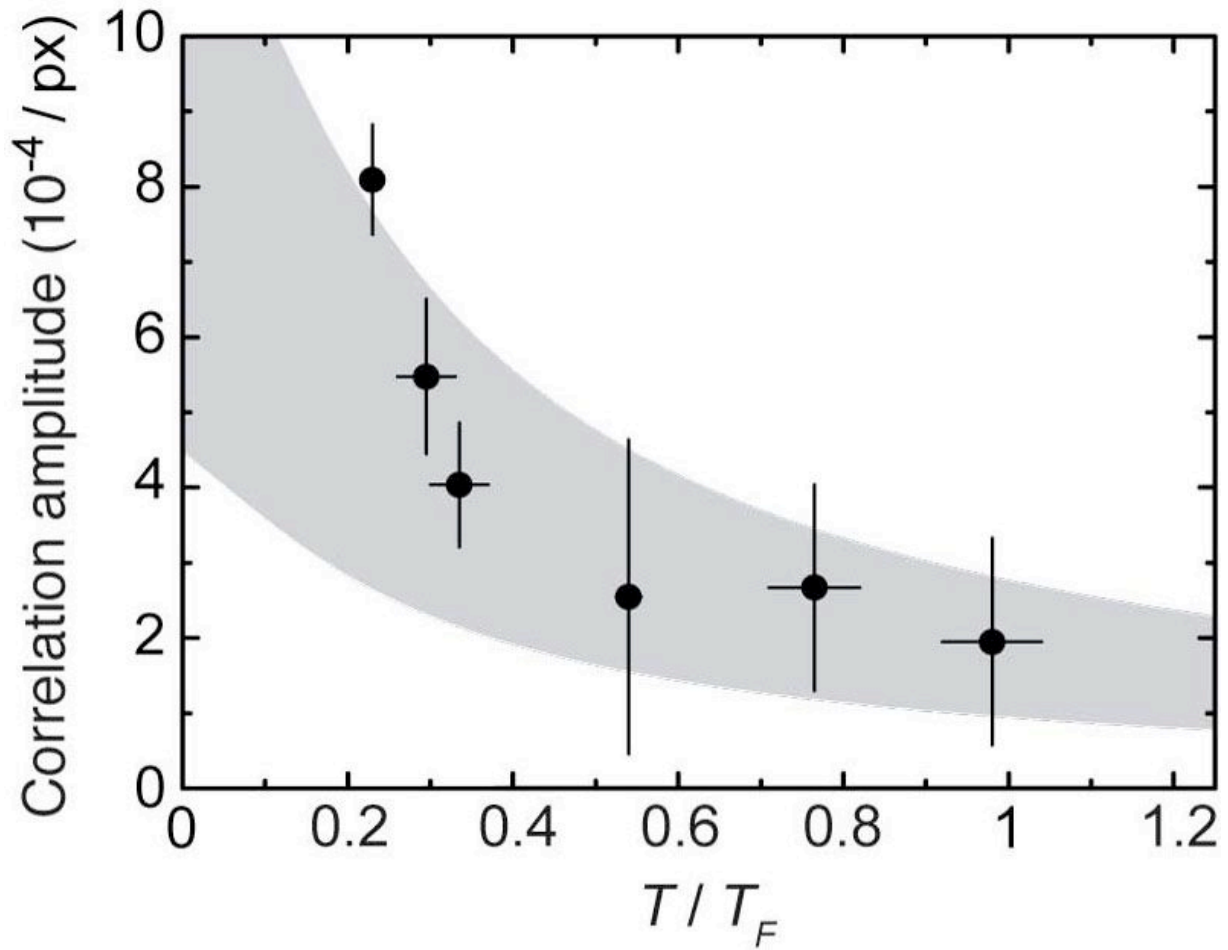

**Figure 3 | Measured correlation amplitude vs. temperature of the atomic cloud before loading into the optical lattice.** For each data point two to four independent runs with about 200 images in total were averaged (error bars 1σ s.d. of the mean). The shaded area indicates the expected values of the correlation amplitude for the two-dimensional version of the theory discussed in the text (see methods), taking into account our experimental uncertainties in the atom number.

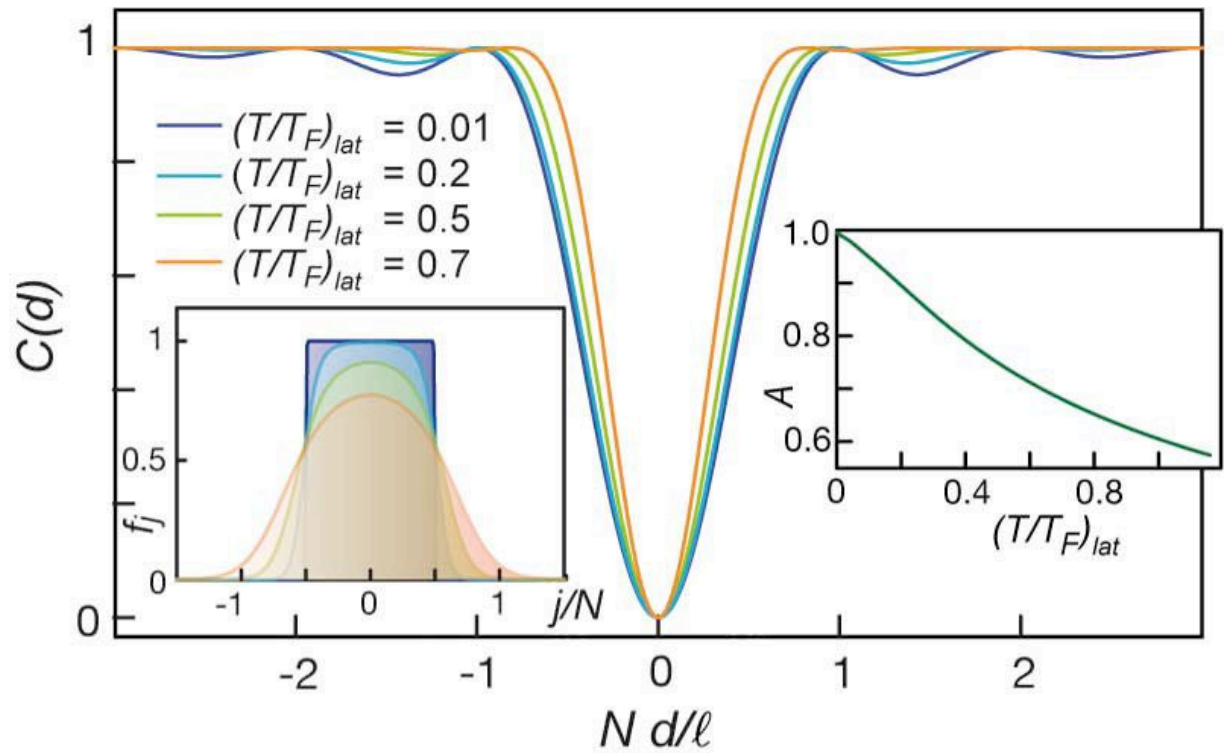

**Figure 4 | Calculated noise correlations for $^{40}$K atoms in a one-dimensional lattice.** The different curves correspond to different values of the temperature $(T/T_F)_{lat}$ in the lattice. The left inset shows the corresponding spatial density profiles of the system for different values of $(T/T_F)_{lat}$. The right inset shows the area $A$ under the correlation dip as a function of $(T/T_F)_{lat}$. The parameters used were $N=100$ and an external confinement of $\omega=2\pi \times 40$ Hz, yielding $T_F^{lat} \approx 350$ nK.